\documentclass[12pt,preprint]{aastex}

\def\logz{\lbrack\hbox{Fe/H}\rbrack}
\def\deg{^{\circ}}

\slugcomment{To appear in the Astrophysical Journal}
\shorttitle{Old Populations of the SMC}
\shortauthors{Dolphin et al.}

\begin{document}

\title{Old Stellar Populations of the Small Magellanic Cloud\linespread{1.0}\footnote{
Based on observations with the NASA/ESA \textit{Hubble Space Telescope},
obtained at the Space Telescope Science Institute, which is operated by
the Association of Universities for Research in Astronomy, Inc., under
NASA contract NAS 5-26555. These observations are associated with proposal
ID 6604.}}
\author{Andrew E. Dolphin\altaffilmark{1}, Alistair R. Walker\altaffilmark{2}, Paul W. Hodge\altaffilmark{3}, Mario Mateo\altaffilmark{4}, Edward W. Olszewski\altaffilmark{5}, Robert A. Schommer\altaffilmark{2}, and Nicholas B. Suntzeff\altaffilmark{2}}

\altaffiltext{1}{Kitt Peak National Observatory, National Optical Astronomy Observatories, P.O. Box 26372, Tucson, AZ 85726; dolphin@noao.edu}
\altaffiltext{2}{Cerro Tololo Inter-American Observatory, National Optical Astronomy Observatories, Casilla 603, La Serena, Chile; awalker@noao.edu, nsuntzeff@noao.edu, rschommer@noao.edu}
\altaffiltext{3}{Astronomy Department, University of Washington, Box 351580, Seattle, WA 98195-1580; hodge@astro.washington.edu}
\altaffiltext{4}{Department of Astronomy, University of Michigan, Ann Arbor, MI 48109; mateo@astro.lsa.umich.edu}
\altaffiltext{5}{Steward Observatory, University of Arizona, Tucson, AZ 85721; edo@as.arizona.edu}

\begin{abstract}
We present WFPC2 and ground-based $VI$ photometry of NGC 121 and a nearby field in the outer SMC.  For NGC 121, we measure a true distance modulus of $\mu_0 = 18.96 \pm 0.04$ (distance of $61.9 \pm 1.1$ kpc), age of $10.6 \pm 0.5$ Gyr, metallicity of $\logz = -1.03 \pm 0.06$, and initial mass of $4.1 \pm 0.4 \times 10^5 M_{\odot}$, assuming a Salpeter IMF with lower cutoff at $0.1 M_{\odot}$.  In the outer SMC field, we find evidence of stars covering a wide range of ages -- from 2 Gyr old to at least $9-12$ Gyr old.  We have measured the distance, extinction, and star formation history (past star formation rates and enrichment history) using a CMD-fitting algorithm.  The distance modulus of the SMC is measured to be $\mu_0 = 18.88 \pm 0.08$, corresponding to a distance of $59.7 \pm 2.2$ kpc.  The overall star formation rate appears to have been relatively constant over this period, although there may be small gaps in the star forming activity too small to be resolved.  The lack of current star-forming activity is a selection effect, as the field was intentionally chosen to avoid recent activity.  The mean metallicity of this field has increased from an average of $\logz = -1.3 \pm 0.3$ for stars older than 8 Gyr to $\logz = -0.7 \pm 0.2$ in the past 3 Gyr.
\end{abstract}

\keywords{galaxies: evolution --- galaxies: stellar content --- Magellanic Clouds}

\section{Introduction}

Studies of the oldest stars of the Magellanic Clouds (MCs) provide an important data point in the study of galaxy formation.  The MCs are sufficiently close that the oldest main sequence turnoff stars ($M_V \sim +4$) can be observed easily through deep ground-based or HST imaging.  In addition, they are distant enough that all stars can be safely assumed (to first order) to be at the same distance, thus avoiding the severe problem one faces when studying the Galactic halo.  The oldest stars of the LMC have been well-studied, with recent work \citep{ols99,hol99,dol00a,har01} indicating two prolonged epochs of star formation, separated by a break in activity from roughly 8 Gyr ago until 4 Gyr ago.  It is likely that the star formation history of the LMC, as well as that of the SMC, has been heavily influenced by past encounters with the Galaxy \citep{mur80}.

The Magellanic Clouds may also hold a key to understanding the formation of our own halo.  Outer halo clusters indicate a wide range of ages and metallicities, suggesting that in the past there has been accretion of several smaller galaxies \citep{sea78}.  That satellite accretion occurs has been illustrated in spectacular fashion by the discovery of the Sagittarius dwarf spheroidal galaxy.  Thus the oldest populations of the Galaxy's companions are likely directly related to the Galaxy's older halo.  However, a definitive comparison will arise only when accurate ages are available for the oldest MC components.  Significant work has been done on the old clusters and field stars of the LMC \citep{ols98,ols99,hol99,dol00a,har01}.  The SMC has been neglected, possibly because of the greater importance of the LMC for distance scale studies, the more complex structure of the SMC \citep{cal86,mat88,hat89}, and its slightly greater distance.

Deep field star population studies for SMC stars of ages greater than a few $10^{8}$ yr began with \citet{gar92}, who covered the entire outer region (r $\geq$ $2\deg$ from the center) with a series of UK Schmidt plates.  From an analysis of the CMDs (which only reached to the level of the HB at $R \sim 20$ mag), they found that the median age in these outer extremities is $\sim 10$ Gyr.  They interpreted the existence of a red horizontal branch in their best CMD as evidence for a population of age 15--16 Gyr which comprises some 7\% of the mass present.  However, the combination of small samples, limited depth, and relatively poor photometric accuracy at faint magnitudes has limited the accuracy of this and other previous ground-based work.

The present work attempts to improve on the previous ground-based work through a combination of HST WFPC2 and ground-based imaging of outer SMC field stars, as well as HST WFPC2 imaging of the oldest cluster in the SMC, NGC 121, with an age of $12 \pm 2$ Gyr \citep{str85}.  We present our data in Section 2, a qualitative examination of the stellar content in Section 3, and a quantitative determination of the star formation history in Section 4.

\section{Observations and Reduction}

\subsection{HST WFPC2 Data}
As part of Hubble Space Telescope program GO-6604, we obtained WFPC2 images of NGC 121 and a nearby SMC field.  The SMC field position was chosen to be well outside the area of recent star formation in the SMC, near NGC 121, but far enough from 47 Tuc to avoid contamination.  Table \ref{tab_fields} gives the coordinates of the five pointings (NGC 121 and four field pointings).

\placetable{tab_fields}

The NGC 121 data consist of four 400s and six 40s images in F555W, and four 500s and four 20s images in F814W.  Guide star acquisition failed during the short F555W images, leaving the telescope guiding on gyroscopes only, but the brevity of the exposures was such that five of the six images (all but dataset U3770509R) were usable.  Additionally, the lack of guide stars prevented HST from returning to same position during subsequent exposures, complicating the cosmic ray removal process.

Cosmic ray removal and photometry was made using the HSTphot package \citep{dol00b}.  The long exposures were taken with two images per orbit, and each pair from the same orbit was combined into cleaned ``deep'' images.  Likewise, the four short F814W images were combined into a cleaned ``short'' image.  Unfortunately, the large guiding errors in the short F555W images (both during and between exposures) prevented the cleaning or combination of those images in any meaningful manner, forcing us to photometer the original uncleaned images.  Figure \ref{fig_ngc121image} shows a cleaned, deep F814W WFPC2 mosaic.

\placefigure{fig_ngc121image}

\begin{figure}
\plotone{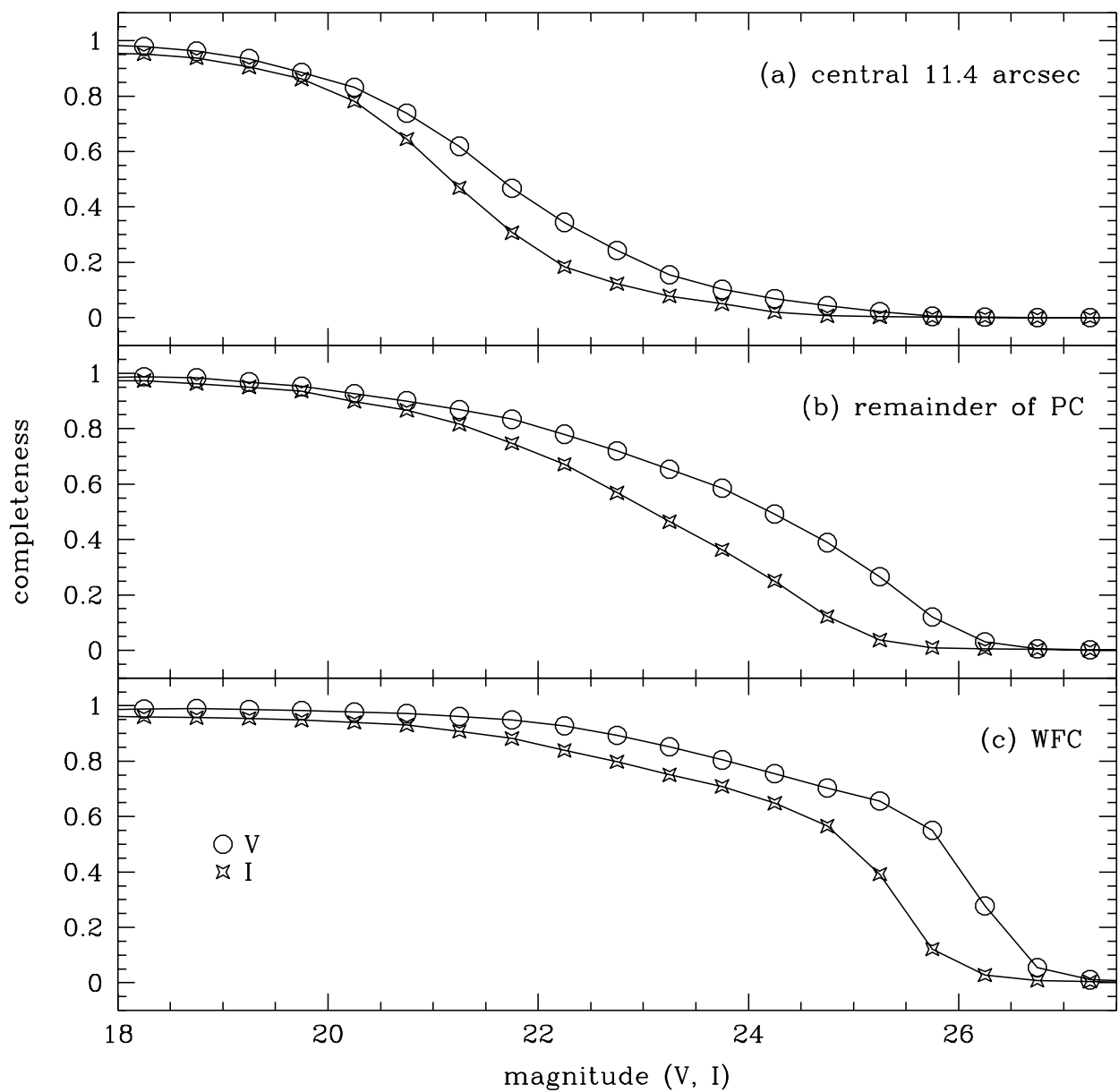}
\caption{$V$ and $I$ completeness of the NGC 121 photometry as a function of magnitude and position, as measured by artificial star tests. Panel (a) shows the completeness function in the inner 11.4 arcsec (250 pixels), panel (b) shows the completeness in the remainder of the planetary camera, and panel (c) shows the completeness in the wide field cameras. Completeness fractions shown are the number of ``good'' stars recovered divided by the number of input stars; the levels never reach 100\% because of bad pixels and the occasional star with bad photometry.}
\label{fig_ngc121comp}
\end{figure}

\begin{figure}
\plotone{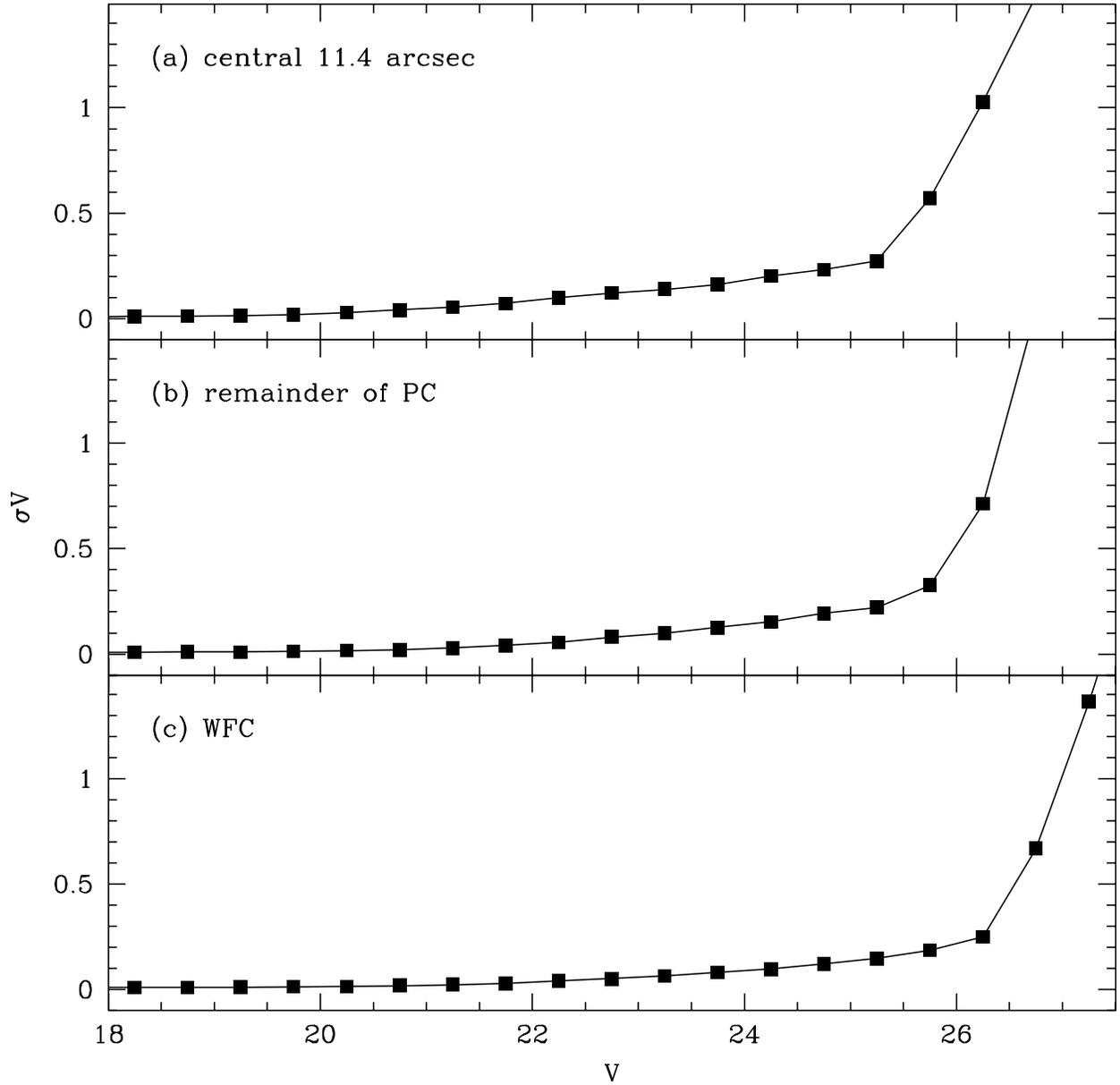}
\caption{$V$ accuracy of the NGC 121 photometry as a function of magnitude and position, as measured by artificial star tests. Panel (a) shows the photometry accuracy in the inner 11.4 arcsec (250 pixels), panel (b) shows the accuracy in the remainder of the planetary camera, and panel (c) shows the accuracy i n the wide field cameras. \label{fig_ngc121phot}}
\end{figure}

The ten resulting images (two deep in each filter, one short in F814W, and five short in F555W) were photometered simultaneously using the HSTphot \textit{multiphot} routine. While \textit{multiphot} was originally written for variable star photometry, its ability to measure magnitudes from images with different pointings, plate scales (F555W vs. F814W), and PSFs (caused by poor guiding in all images) made it an ideal choice for these data.  We also ran artificial star tests, with artificial stars distributed in position similarly to the image and evenly on a color-magnitude diagram (CMD).  Figure \ref{fig_ngc121cmd} shows our full CMD (all stars) and our clean CMD (only stars with good photometry -- $\chi \le 2.0$ and $|\mbox{sharpness}| \le 0.3$).  Because of a strong dependence of completeness on position and the desire to use artificial star tests to later reconstruct the CMD, we generated a large number (over 1.2 million) of artificial stars and stored the results (completeness fraction and distribution of output minus input magnitudes and colors) as functions of position and input photometry in an artificial star database.  Figure \ref{fig_ngc121comp} shows the completeness functions at various distances from the center of NGC 121, while Figure \ref{fig_ngc121phot} shows the photometry accuracy as a function of position, as judged from output minus input $V$ magnitudes of the artificial star tests.

\placefigure{fig_ngc121cmd}
\placefigure{fig_ngc121comp}
\placefigure{fig_ngc121phot}

\begin{figure}
\plotone{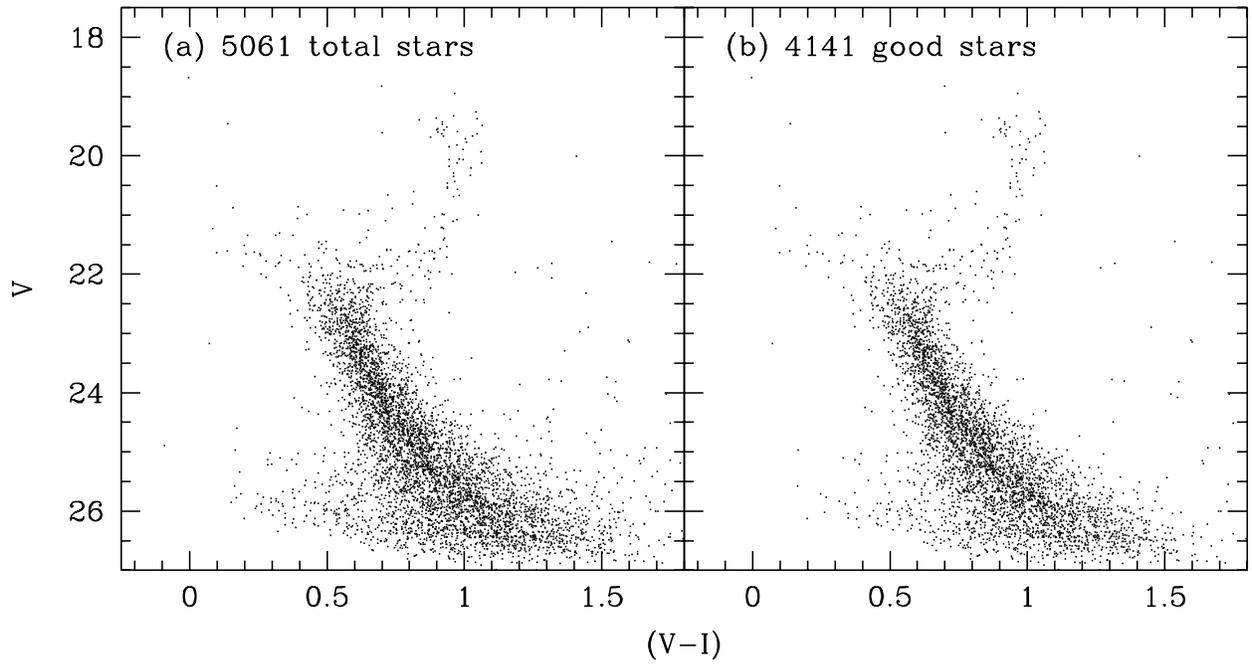}
\caption{WFPC2 color-magnitude diagram of the SMC field. Panel (a) shows all stars located by \textit{multiphot}, while panel (b) shows just the stars with good photometry ($\chi \le 2.0$ and $|\mbox{sharpness}| \le 0.3$).}
\label{fig_fieldcmd}
\end{figure}

\begin{figure}
\plotone{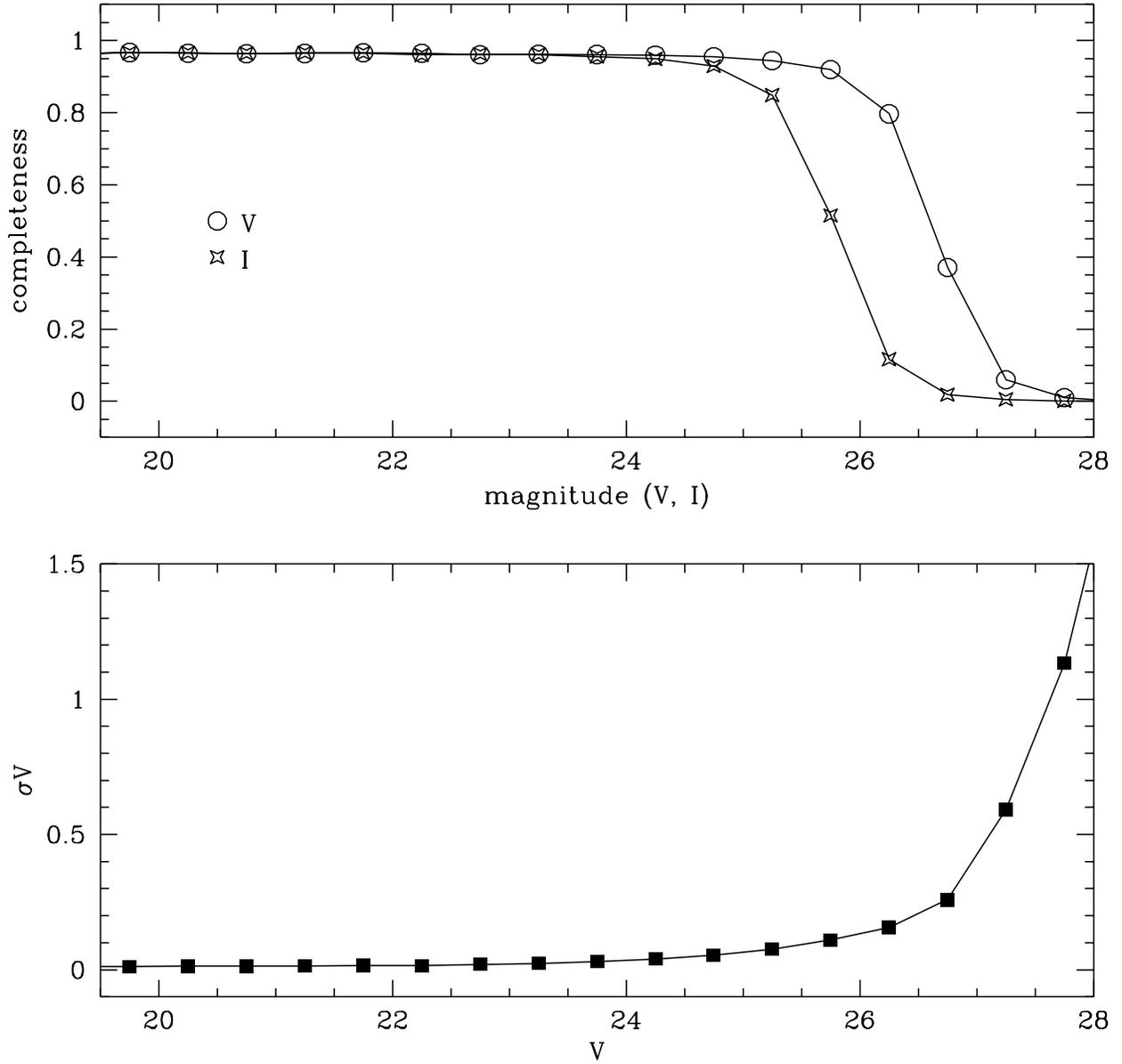}
\caption{$V$ and $I$ completeness and $V$ accuracy of the SMC field photometry as functions of magnitude from the WFPC2 data.}
\label{fig_fieldcomp}
\end{figure}

Photometry of the SMC field was much more straightforward.  The data consist of four 500s F555W and four 600s F814W exposures at each of four pointings, separated by $\sim6-7$ arcmin.  There were no significant guiding problems, allowing us to generate cleaned 2000s F555W and 2400s F814W images at each pointing.  These were also photometered with \textit{multiphot}.  Because of the small number of bright stars usable for aperture corrections, we are only using the WFC data from these pointings.  As with NGC 121, we constructed a database of artificial star tests.  However, as the crowding was essentially constant over the images, we ran only 375,000 artificial stars, distributed evenly across the images and the CMD.  Figure \ref{fig_fieldcmd} shows our resulting full and clean CMDs, and Figure \ref{fig_fieldcomp} shows the completeness and photometry accuracy.

\placefigure{fig_fieldcmd}
\placefigure{fig_fieldcomp}

\begin{figure}
\plotone{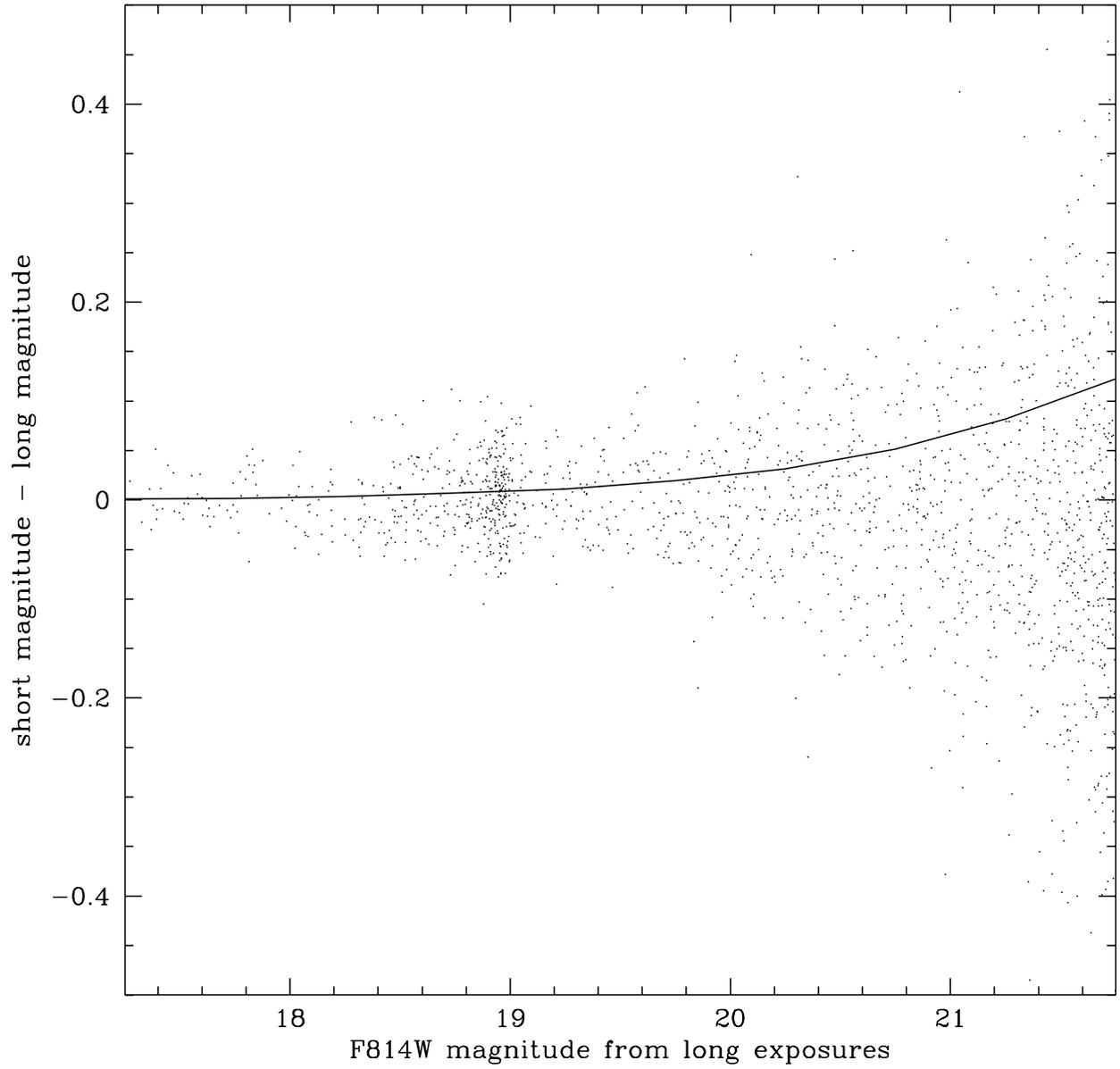}
\caption{Comparison of F814W magnitudes from short (20s) and long (500s) exposures after CTE corrections. The solid line is the correction suggested by \citet{cas98}; however, our data show no significant differences.}
\label{fig_longshort}
\end{figure}

We used the CTE correction and calibrations given by \citet{dol00c} to transform our photometry to the standard $VI$ system.  Because we have both long and short exposures, these data allow the opportunity to test for the presence of the so-called ``long vs. short'' anomaly.  Figure \ref{fig_longshort} shows a comparison of mean F814W magnitudes from our short (20s) images and long (500s) images, along with the trend found by \citet{cas98}.  As our data show no measurable systematic differences, we will follow the suggestion of \citet{dol00c} and make no correction.

\placefigure{fig_longshort}

\subsection{Ground-based Data}
Ground-based photometry of outer SMC field stars was obtained using the CTIO 4m telescope and prime focus (PFCCD) camera, and with the 0.9m telescope and Cassegrain focus camera (CFCCD).  Both cameras utilized SITe 2048 CCDs, read through four amplifiers with ARCON controllers.  The field position was chosen to be $2.0\deg$ NE of NGC 121; a position closer to NGC 121 would have increased the contamination from the Galactic globular cluster 47 Tuc.  The region we have chosen is well away from the disturbed ``wing'' and ``bridge'' regions of the SMC, and shows no indication of significant line-of-sight depth.

The 4m observations were made on the nights of 20-23 November 1995.  Two of these nights were fully photometric and a third night was photometric during the time the SMC observations were made.  Image FWHM varied between 0.8 and 1.2 arcsec.  Exposures (typically $10 \times 300$s $V$, $10 \times 300$s $I$) were made in each of four $15 \times 15$ arcmin fields centered on the HST SMC pointings.  The  telescope was dithered a few arcseconds between exposures to aid in flat-fielding.  During photometric conditions exposures were taken of \citet{lan92} standard fields, and supplementary $I$ band exposures of an almost star-free field were taken in order to prepare a fringe frame; this in combination with twilight sky flat fields reduced systematics to well below one percent.

The 0.9m observations were made on the night of 8 November 1999, and consisted of short (100s $V$, 60s $I$) and long (300s $V$, 200s $I$) exposures for each field, $13.6 \times 13.6$ arcmin in size.  The night was photometric, and measurements were made of 105 \citet{lan92} standards in ten fields over a wide range in airmass; 86 of these stars were retained in the final color equation fits. Twilight sky flat field exposures were used to correct systematics to well below one percent.

The 0.9m standard star observations confirmed the good quality of the night; the uncertainties of the $V$ and $V-I$ zeropoints were each 0.002 mags and the extinction coefficients were 0.13 ($V$) and 0.05 ($I$). Stars on the SMC frames were located and photometered using DAOPHOT \citep{ste87}, then transformed to the $VI$ system by using the same aperture photometry parameters as had been used for the standard star measurements.  All stars on the 0.9m SMC frames were subsequently measured with PSF-fitting; strong coma across the 0.9m field necessitates use of a PSF varying cubically with field position.

\begin{figure}
\plotone{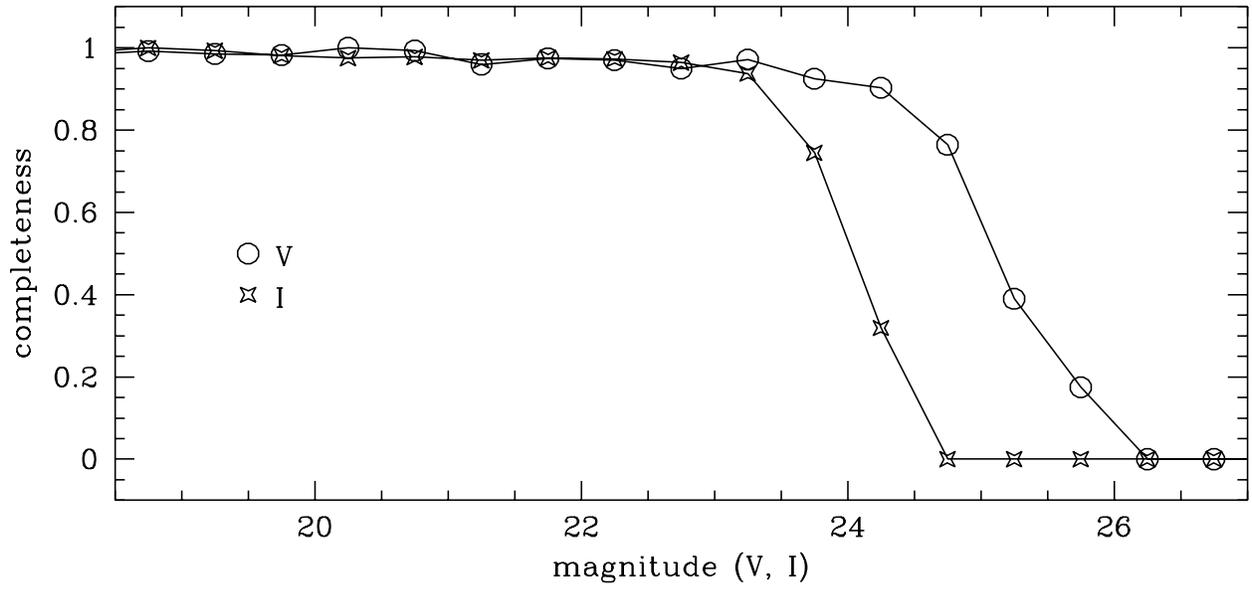}
\caption{$V$ and $I$ completeness as functions of magnitude from the ground-based data.}
\label{fig_fieldcomp2}
\end{figure}

The 4m data set, also measured using DAOPHOT, was calibrated with reference to the 0.9m data.  The accuracy of the zeropoints is thought to be no worse than $\pm 0.01$ mag in both $V$ and $V-I$, as ascertained by combining in quadrature errors resulting from the various steps in proceeding from the standard star photometry to the SMC.  Figure \ref{fig_fieldcmd2} shows our resulting CMD, and Figure \ref{fig_fieldcomp2} shows the completeness.  Our photometry will be made available at ADC.

\placefigure{fig_fieldcmd2}
\placefigure{fig_fieldcomp2}

\section{Distance and Stellar Content}

\subsection{NGC 121 \label{sec_121pop}}

We first turn our attention to NGC 121.  The CMD shown in Figure \ref{fig_ngc121cmd} shows very sharp features, indicating that (as one would expect for a globular cluster) NGC 121 is a single-population object.  The clean CMD in panel (b) shows only the expected globular cluster populations (main sequence, red giant branch, horizontal branch, and asymptotic giant branch) and a few foreground and background stars.  The primary source of foreground contamination in NGC 121 is 47 Tuc, whose main sequence crosses the NGC 121 RGB at about the level of the NGC 121 horizontal branch.  However, although such stars are clearly present in the NGC 121 data, it is not a major source of contamination.  We note that some of the apparent scatter in the horizontal branch region is also due to the presence of RR Lyraes.  More problematic is the SMC background, which contributes the stars seen above the NGC 121 main sequence turnoff and gives the impression of more blue stragglers than are really there.  A comparison of Figures \ref{fig_ngc121cmd} and \ref{fig_fieldcmd} shows that at least half of the apparent blue stragglers seen in the NGC 121 field, and nearly all of those in the WFC chips, are SMC field main sequence field stars.

The distance to NGC 121 can be measured through a variety of methods.  The $I$ magnitude of the RGB tip \citep{dac90,lee93} is relatively insensitive to age and metallicity, and thus can serve as a standard candle.  However, the exact position of the RGB tip in this CMD is extremely uncertain due to the scarcity of stars.  The brightest star in the CMD falls at $I = 14.34$, which if it were truly an RGB star would bring the SMC closer than the LMC.  Adopting the second-brightest star ($I = 14.60$) and the top of the well-populated RGB ($I=15.31$) as our limits, we estimate the RGB tip position at $I = 15.0 \pm 0.4$.  Adopting an extinction of $A_V = 0.10$ from the \citet{sch98} maps and an RGB tip absolute magnitude of $M_I = -4.05$ calculated from the \citet{lee93} calibration, we find a true distance modulus of NGC 121 of $\mu_0 = 19.0 \pm 0.4$.

Given the quality the data presented in this paper, a more accurate measurement would involve the horizontal branch.  We measure a horizontal branch mean magnitude of $V = 19.69 \pm 0.03$.  Again adopting an extinction of $A_V = 0.10 \pm 0.03$ and using the absolute magnitude calibration of $M_V = 0.61 \pm 0.09$ \citep{car00}, we find a distance modulus of NGC 121 of $\mu_0 = 18.98 \pm 0.10$.

The metallicity can be measured photometrically, using the color of the RGB at $M_I = -3.5$ \citep{lee93}.  For our NGC 121 CMD, we find $(V-I) = 1.52 \pm 0.02$ or $(V-I)_0 = 1.48 \pm 0.03$ at this position on the RGB, which corresponds to a metallicity of $\logz = -1.22 \pm 0.08$.  This metallicity corresponds to $\logz = -1.02$ on the scale given by \citet{car97}, which we will adopt because of its greater consistency (in terms of predicted RGB colors) with theoretical models.

As noted above, this is the oldest globular cluster in the SMC and the only one containing RR Lyraes \citep{wal89}.  However, its CMD shows a very red horizontal branch, indicating that it is likely much younger than Galactic and LMC globular clusters.  This can be tested using the $\Delta V^{TO}_{HB}$ age measurement \citep{cha96}. We measure the turnoff position to be $V = 22.98 \pm 0.05$, giving $\Delta V^{TO}_{HB} = 3.29 \pm 0.06$.  Adopting the calibrations given by \citet{cha96} that are closest to the horizontal branch $M_V$ value used here, we find an age of $t = 9.4 \pm 0.9$ Gyr (on this scale, the age of 47 Tuc is 12.7 Gyr).

\begin{figure}
\plotone{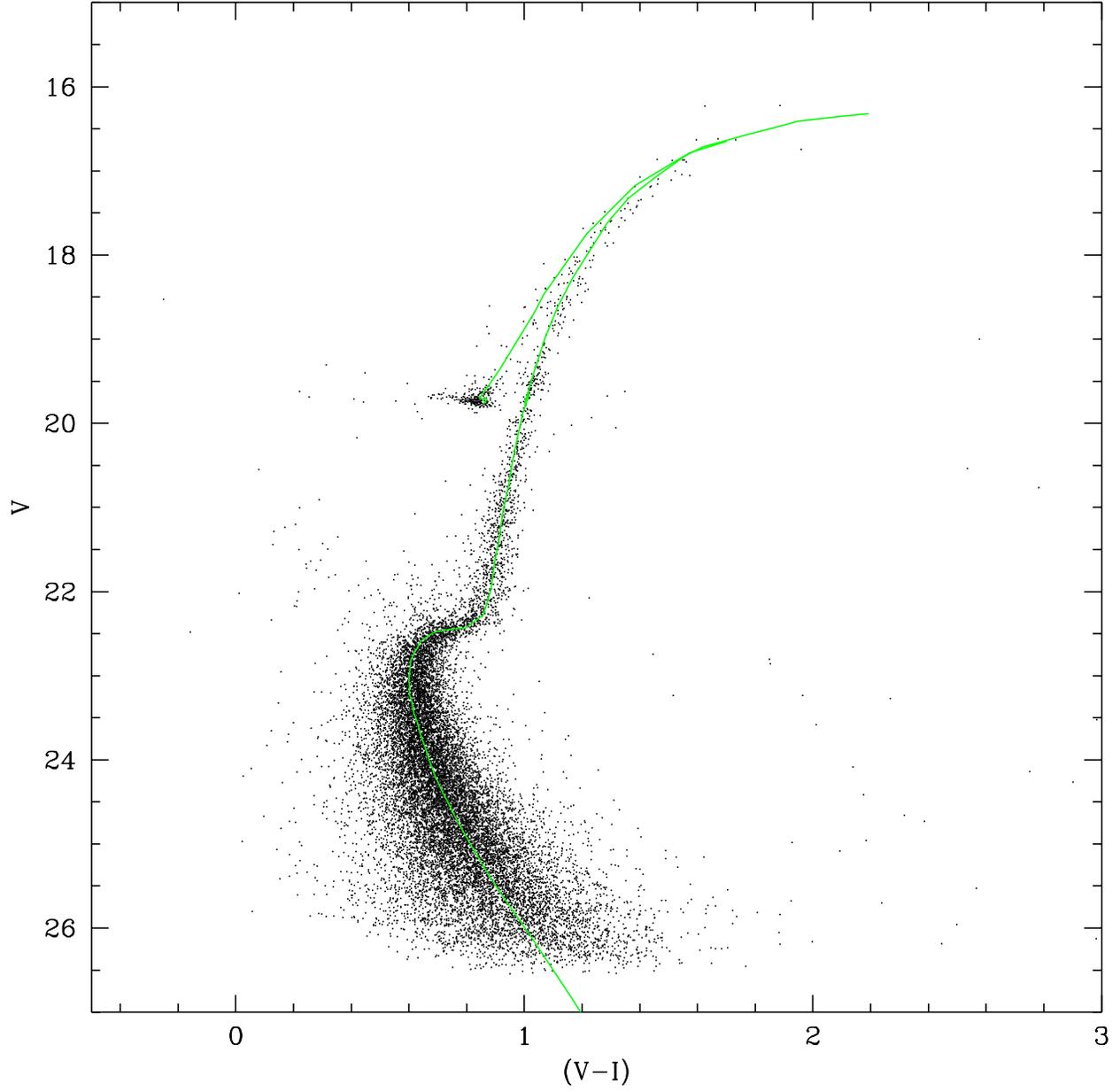}
\caption{NGC 121 CMD, with best-fitting isochrone overplotted. The parameters are $\mu_0 = 18.96$, $A_V = 0.04$, $t = 10.6$ Gyr, and $\logz = -1.03$.}
\label{fig_ngc121iso}
\end{figure}

We can also apply a CMD reconstruction technique similar to that described below, in Section \ref{sec_sfh}, to attempt to measure the properties of NGC 121.  Such a solution has several advantages -- most notably, we obtain a single self-consistent solution (distance, extinction, age, and metallicity) in one step.  This also provides an independent check on the values given above, as well as testing the accuracy of our CMD reconstruction technique.  To make this solution, we divided the CMD into 0.1 magnitude ($V$) by 0.05 magnitude ($V-I$) bins, counting the number of stars in each.  We then generated synthetic CMDs using \citet{gir00} isochrones, our artificial star library, and the field star CMD as a background CMD.  By making a large number of synthetic CMDs, each with a different combination of distance, extinction, age, and metallicity, we are able to directly solve for the four parameters through a comparison of the observed and synthetic CMDs.  The description of our fit parameter and derivation of uncertainties is described in Section \ref{sec_sfh}.  The calculated parameters from this solution are $\mu_0 = 18.96 \pm 0.03$, $A_V = 0.04 \pm 0.03$, $t = 10.6 \pm 0.5$ Gyr, and $\logz = -1.03 \pm 0.06$ (random errors only).  We also measured an initial cluster mass of $3.25 \pm 0.09 \times 10^5 M_{\odot}$ for the observed portion of the cluster (79\% of the total cluster), assuming a Salpeter IMF and mass limits of 0.1 and 120 $M_{\odot}$.  These values are all consistent with those calculated above, and we will adopt these because of the smaller uncertainties that result from considering the entire CMD rather than just the positions of three features (HB magnitude, RGB color, and MSTO magnitude).  Figure \ref{fig_ngc121iso} shows the NGC 121 CMD (from Figure \ref{fig_ngc121cmd}b) with this isochrone overplotted.

\placefigure{fig_ngc121iso}

\subsection{Field Stars}

\begin{figure}
\plotone{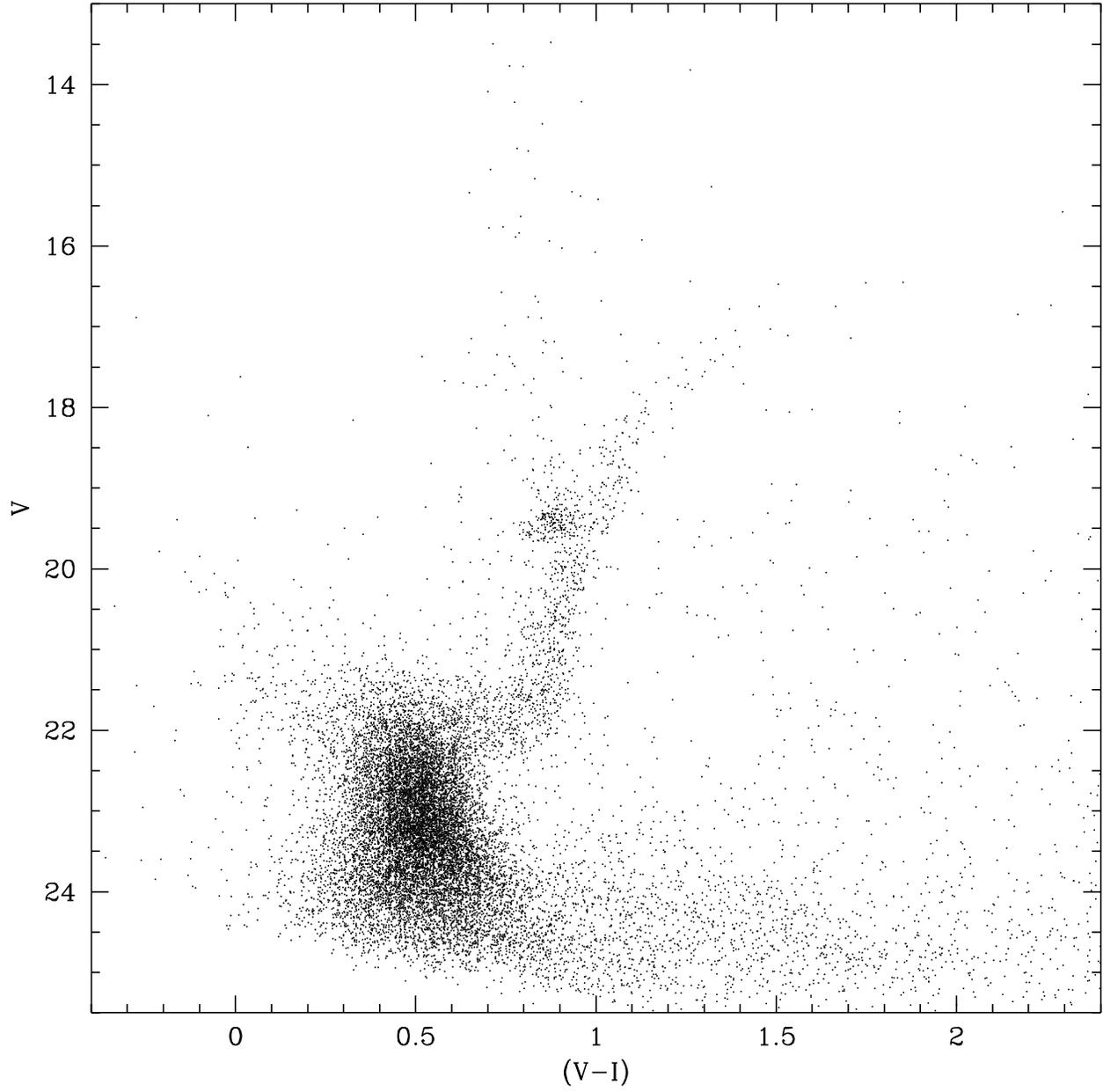}
\caption{Ground-based color-magnitude diagram of the SMC field.}
\label{fig_fieldcmd2}
\end{figure}

\begin{figure}
\plotone{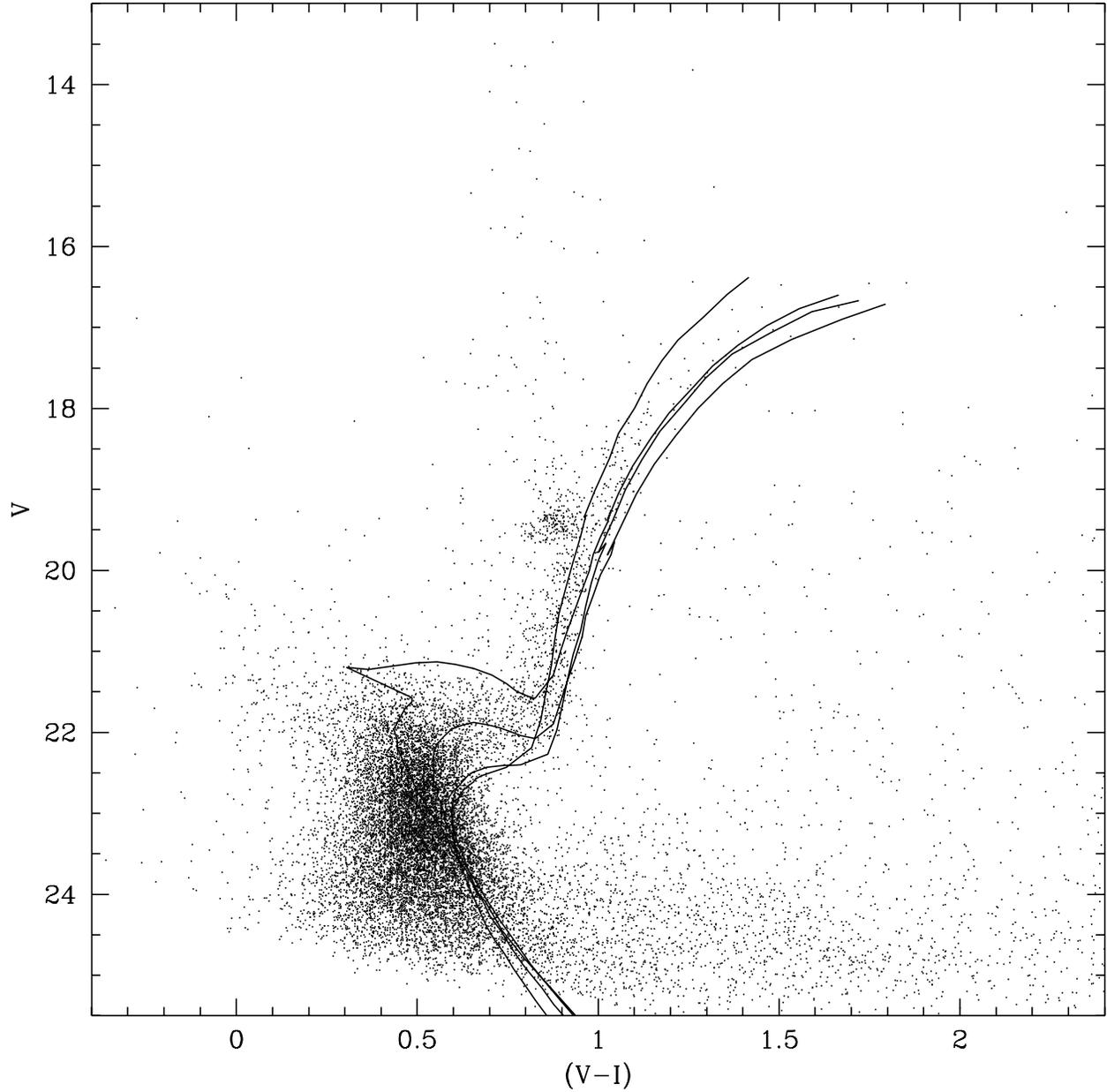}
\caption{SMC field CMD, with isochrones overplotted. From youngest to oldest, isochrones are 2.5 Gyr, $\logz = -0.7$; 5 Gyr, $\logz = -0.7$; 10 Gyr, $\logz = -1.0$; and 15 Gyr, $\logz = -1.5$. Isochrones are taken from the models of \citet{gir00}, and are plotted with $\mu_0 = 18.96$ and $A_V = 0.04$.}
\label{fig_fieldiso}
\end{figure}

In contrast to the sharp, well-defined CMD observed for NGC 121, the ground-based field star CMDs (Figures \ref{fig_fieldcmd} and \ref{fig_fieldcmd2}) show a prolonged star formation history.  Figure \ref{fig_fieldiso} shows the field CMD, with isochrones of various ages from 2.5 to 15 Gyr overplotted (using the distance and extinction found above for NGC 121).  There is a tight red clump, centered at $I = 18.56 \pm 0.03$ and $(V-I) = 0.88 \pm 0.03$, and a few possible horizontal branch stars, indicators of intermediate-aged and ancient populations, respectively.  The clump magnitude is significantly fainter than the value of $I_0 = 18.33 \pm 0.05$ reported by \citet{uda98}, implying significant population differences between the fields examined -- fields closer to the center have much more recent star formation and subsequently a brighter clump, as suggested by \citet{gir01}.  The RGB is also well-populated, but falls off above the red clump because such stars would have been saturated in our data (we only obtained long images at the field pointings).  The data also contain a weak young ($< 2$ Gyr) main sequence population.

\placefigure{fig_fieldiso}

The bulk of our information regarding the star formation history comes from the turnoff region.  Judging from the large spread in this region (compared with the sharp turnoff of NGC 121), along with the presence of populations ranging from a young main sequence to an old red clump, it seems clear that the SMC has been forming stars over much of its history.  The relatively even spread of stars in the turnoff region implies that the star formation history has not been highly bursty, as strong bursts of star formation separated by long quiescent periods would create a color-magnitude diagram with multiple discrete turnoffs.  We will attempt to quantify the star formation history in the next section.

\begin{figure}
\plotone{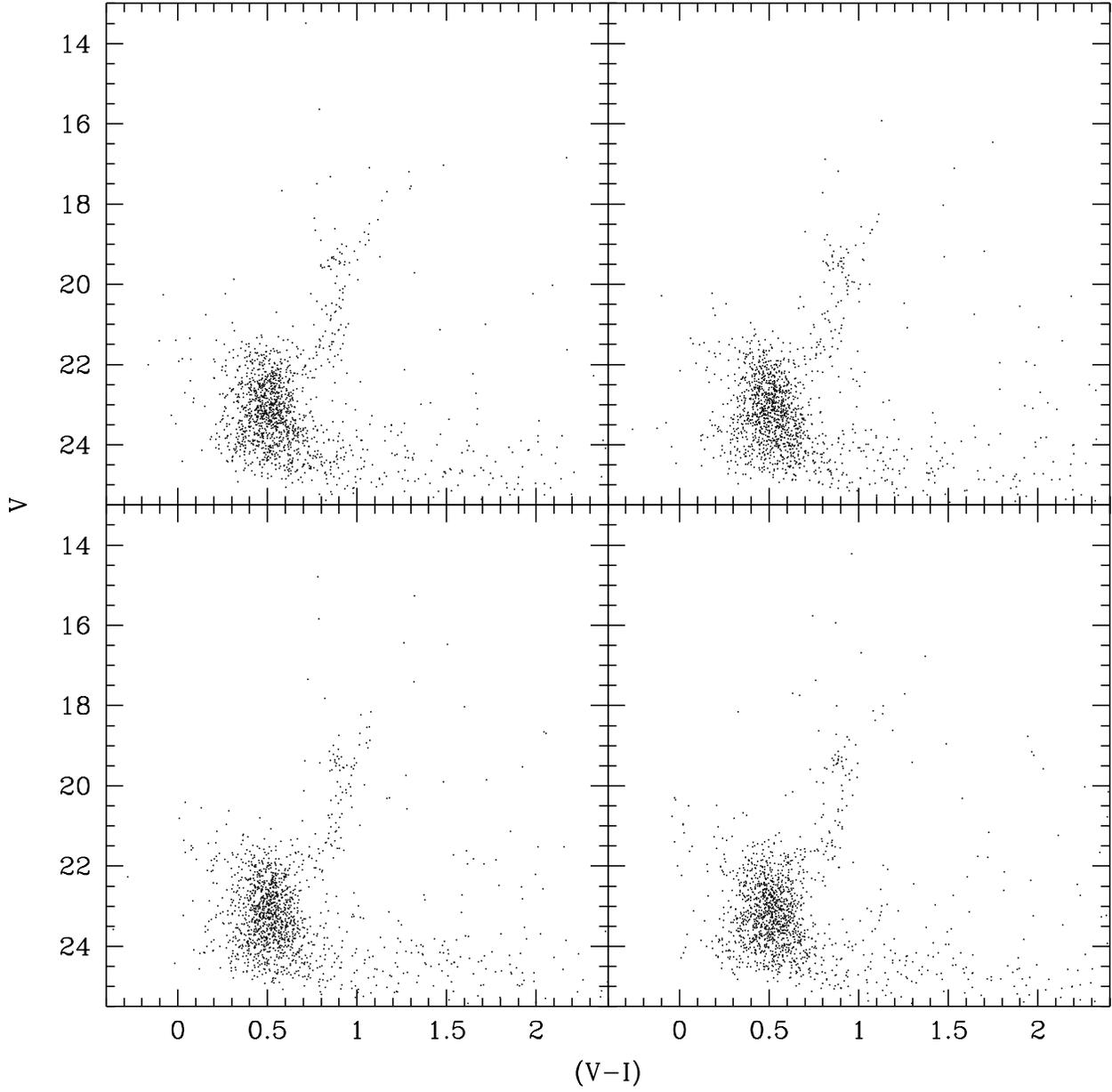}
\caption{Four random samplings of the ground-based CMD, each with 9\% of the stars.}
\label{fig_fieldcmd_test}
\end{figure}

A comparison between Figures \ref{fig_fieldcmd} (our WFPC2 data) and \ref{fig_fieldcmd2} (our ground-based data) shows the strengths of each data set.  The CMD from the WFPC2 data is extremely clean (very few stars falling in non-physical parts of the CMD) due to HST's superior resolution.  However, the smaller field of view produces over an order of magnitude fewer evolved stars, magnifying confusion from Poisson noise.  Specifically, we note that the isochrones overlaid in Figure \ref{fig_fieldiso} fall at the positions of apparent bursts in the WFPC2 data; the lack of such structure in the main sequence turnoff of the ground-based CMD implies that the apparent gaps in the WFPC2 CMD are not real, but are caused by the small-number statistics.  To quantify this uncertainty, we show four random samplings of the ground-based CMD in Figure \ref{fig_fieldcmd_test}, each containing 9\% of the stars to simulate the combined size of our four WFPC2 fields.  While Figure \ref{fig_fieldcmd2} shows a broad and smooth main sequence turnoff, three of the four panels in Figure \ref{fig_fieldcmd_test} show what appear to be gaps in the turnoff region, which would likely be interpreted as gaps in the star formation history.  Thus, while the WFPC2 CMD will be used below in our distance, extinction, and metallicity measurement (because of its superior photometry quality), the star formation history will be entirely based on the ground-based CMD.

\placefigure{fig_fieldcmd_test}

\section{Star Formation History \label{sec_sfh}}

While the previous section described the star formation history of the SMC field rather qualitatively, we can attempt to make a quantitative measurement through CMD reconstruction.  The method used in this paper was first proposed by \citet{dol97}, and the present form of the procedure is described by \citet{dol01}, and the reader is referred to those papers for detailed descriptions.

The goal of this procedure is to determine the star formation history that produces a synthetic CMD most resembling the observed CMD.  While the NGC 121 solution described in Section \ref{sec_121pop} applied the constraint of a single-population object, this is not possible for field stars, as the CMD (as noted in the previous section) shows stars with a wide range of ages and/or metallicities.  Instead, for an assumed of distance and extinction, we generate a library of synthetic CMDs covering the needed age and metallicity ranges, each synthetic CMD created by convolving a theoretical isochrone from the \citet{gir00} models with our artificial star library.  By determining the combination of synthetic CMDs that best matches the observed data, we are able to measure the star formation history.  This procedure is repeated at a range of distance and extinction values, and those producing good fits to the observed CMD are used for calculating the distance, extinction, and star formation rate (a function of time and metallicity), as well as uncertainties for all values.  Finally, we test the quality of the fit by Monte Carlo tests.

\begin{figure}
\plotone{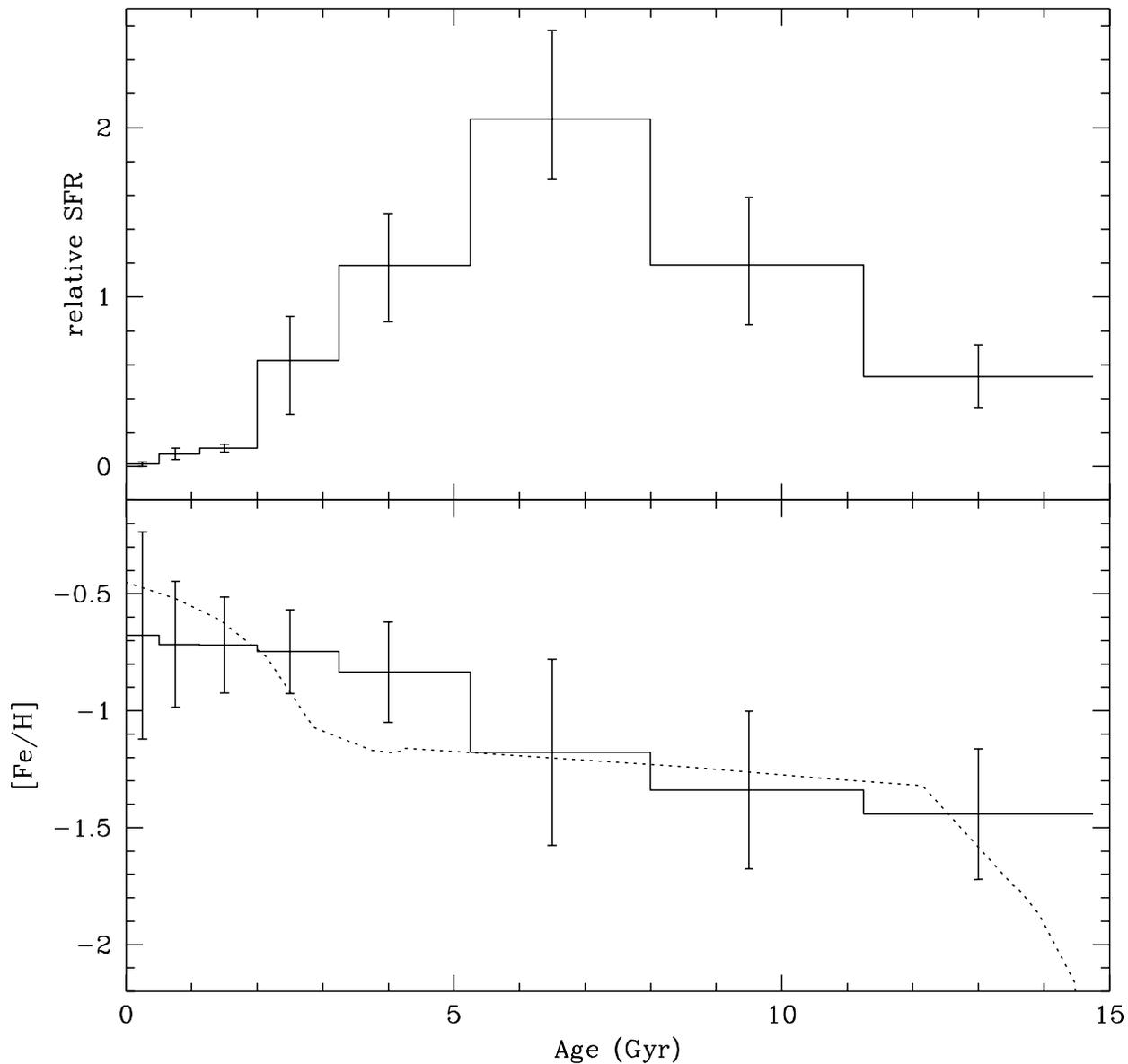}
\caption{Calculated star formation history of the SMC field. The top panel shows the recovered relative star formation rates; the bottom panel shows the enrichment history. Note that both panels show only the mean values of star formation rate and metallicity within each age range -- any short bursts would be averaged out. For comparison, the age-metallicity relation derived by \citet{pag98} is shown with the dotted line.}
\label{fig_fieldsfh}
\end{figure}

\begin{figure}
\plotone{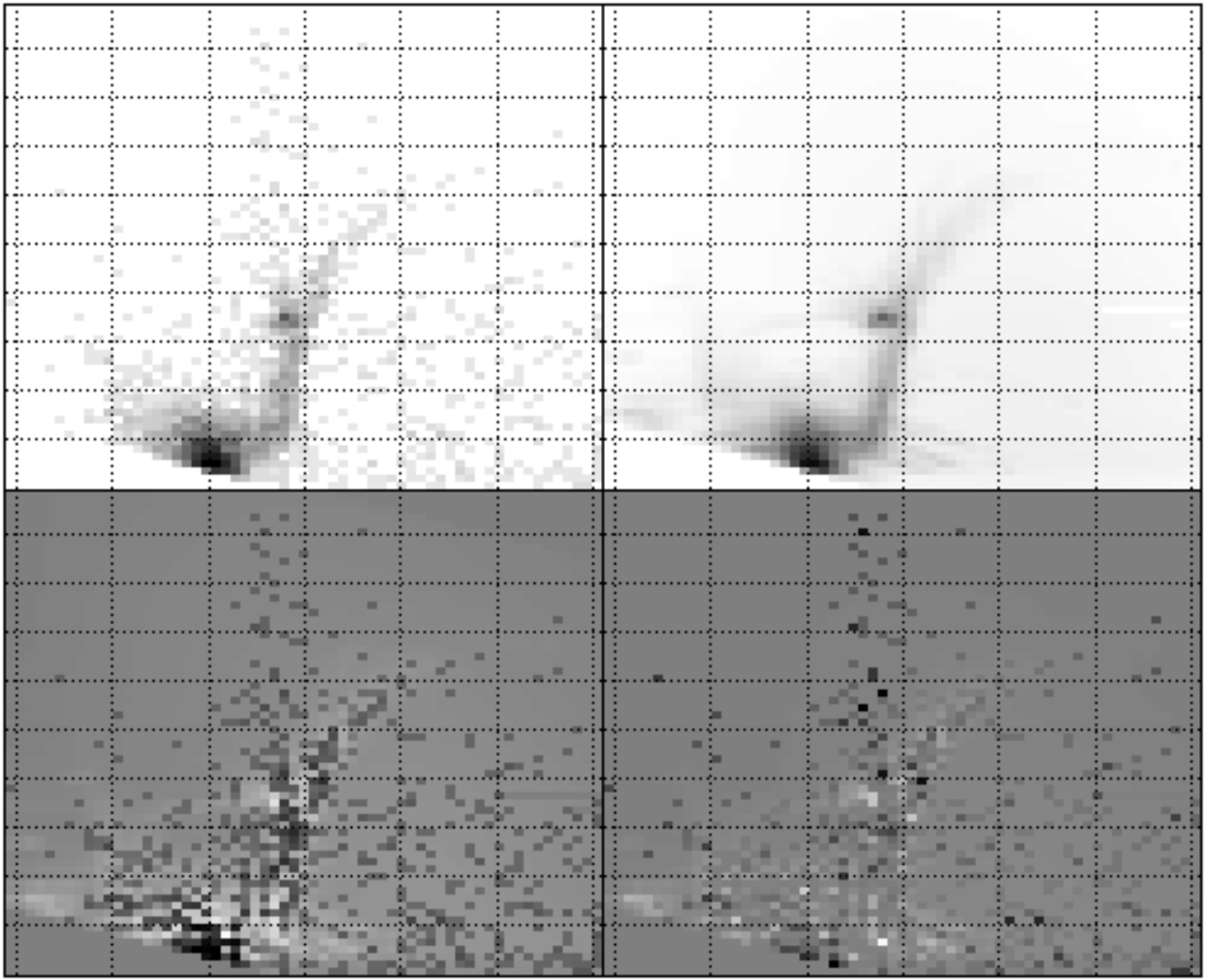}
\caption{Observed and recovered color-magnitude diagrams of the ground-based SMC field.  Upper-left is the observed CMD, binned into 0.05 magnitude ($V-I$) by 0.15 magnitude ($V$) bins.  The synthetic CMD of the best fit is shown, plotted identically, in the upper-right panel.  The observed minus synthetic CMD residual, magnified in color contrast by a factor of five, is shown in the lower-left.  Finally, the lower-right panel shows the significance of the residual differences, with the white and black points corresponding roughly to $2 \sigma$ errors.  In all panels, the CMD limits are $13 < V < 23$ and $-0.5 < V-I < 2.5$, with the grid plotted every magnitude in $V$ and half magnitude in $V-I$.}
\label{fig_fieldhess}
\end{figure}

From these data, we find the star formation history given in Table \ref{tab_fieldsfh} and plotted in Figure \ref{fig_fieldsfh}.  The color-magnitude diagram derived from this fit is shown in Figure \ref{fig_fieldhess}.  The quality of the fit was mediocre, with an overall effective $\chi^2$ of 2.1.  However, the largest contribution to this error comes from the foreground stars and the stripes in the model CMD (produced by an insufficient number of artificial stars).  The measured parameters of $\mu_0 = 18.86 \pm 0.10$ and $A_V = 0.01 \pm 0.12$ were both consistent with those determined for NGC 121.  The cleaner WFPC2 data could be fit better, with an overall effective $\chi^2$ of 1.5.  The measured distance and extinction were $\mu_0 = 18.89 \pm 0.12$ and $A_V = 0.09 \pm 0.10$.  Because of minimal overlap between the ground-based and WFPC2 fields (the ground-based data included only one of the four WFPC2 fields), we combine these mostly-independent measurements to obtain values of $\mu_0 = 18.88 \pm 0.08$ and $A_V = 0.05 \pm 0.07$ for this region of the SMC.  As with the CMD-based measurements for NGC 121, these uncertainties account for photometric errors and shot noise in the CMDs, but not for possible systematic errors in the models.

\placetable{tab_fieldsfh}
\placefigure{fig_fieldsfh}
\placefigure{fig_fieldhess}

Examining the past star formation rates, which are based only on the ground-based data, we find what appears to be a broadly peaked star formation history, with the largest star formation rate between 5 and 8 Gyr ago.  It is also consistent at the $2 \sigma$ level with a constant star formation rate from 15 Gyr ago until $\sim 2$ Gyr ago.  Assuming a Salpeter IMF from 0.1 to 120 $M_{\odot}$, the mean star formation rate within this field is $2.1 \pm 0.1 \times 10^{-5} M_{\odot} yr^{-1}$, corresponding to a star formation intensity of $3.2 \pm 0.2 \times 10^{-4} M_{\odot} kpc^{-2} yr^{-1}$.  We have made the same calculations using the WFPC2 field star photometry, and measured a consistent star formation history and overall star formation rate, but the uncertainties are extremely large due to the much smaller field of view.  Our measured star formation rates are consistent with the findings of \citet{gar92}, who found that $\sim 7$\% of the stars were 15 Gyr old (we measure $14 \pm 5$\% of the star formation before 11 Gyr ago), that the mean stellar age is $\sim 10$ Gyr in the outer regions of the SMC (we measure a mean age of 7.5 Gyr).  We would also agree with \citet{har84}, who commented that the SMC likely contains stars of all ages.  After this long period of continuous star formation activity, the star formation rate in this location appears to have dropped by an order of magnitude $\sim 2$ Gyr ago and stopped entirely in the past 0.5 Gyr.

An interesting question is that of the age of the oldest stars in this region of the SMC.  In our star formation history solutions above, we assumed a canonical maximum age of 15 Gyr.  However, the 11-15 Gyr bin was only weakly detected, leaving the question of whether or not such stars exist in measurable amounts in this field.  As demonstrated in Figure \ref{fig_fieldiso}, the age-metallicity pseudo-degeneracy would permit one to ``hide'' a metal-poor 15 Gyr population in the CMD, with the only unambiguous trace being a handful of blue horizontal branch stars that are observationally indistinguishable from younger stars, foreground contamination, binaries, or blends.  So while it is impossible to rule out the existence of 15 Gyr old stars, we can conduct a test for the oldest measurable population by fitting the star formation history with a variety of maximum ages.  We find that excellent fits (equally good as our fit using 15 Gyr) can be obtained with maximum ages as young as 12 Gyr, and that acceptable fits can be obtained with maximum ages as young as 9 Gyr.  This indicates that there are stars at least as old as 9 Gyr in the field and probably stars as old as 12 Gyr, but we cannot rule out the possibility of small amounts of older stars.  With the maximum age set to 9 Gyr, we measure an identical star formation history to that in Table \ref{tab_fieldsfh}, except that all star formation older than 6 Gyr has been compressed into a 3 Gyr range to create a strong initial burst of star formation activity rather than the gradual increase to maximum seen in Table \ref{tab_fieldsfh}.  This age constraint of $9-12$ Gyr is similar to that indicated by the presence of field RR Lyraes in this region of the SMC \citep{gra75}.  Adopting NGC 121 as the youngest accurately-measured globular cluster with RR Lyraes (age = $10.6 \pm 0.5$ Gyr) and allowing for metallicity effects, we likewise find that there must be stars with ages of at least $\sim 10$ Gyr in the field population.

In terms of chemical enrichment, we first note that no assumptions were made regarding the SMC chemical enrichment history.  The values in Table \ref{tab_fieldsfh} are entirely from the numerical star formation history solutions, and we find it encouraging that the solution passes the ``sanity test'' of finding metallicity increasing with age, from $\logz = -1.3 \pm 0.3$ at ages greater than 8 Gyr (or $\logz = -1.4 \pm 0.2$ at ages greater than 11 Gyr, if older stars are present) to $\logz = -0.7 \pm 0.2$ for stars younger than 3 Gyr.  We also note that the metallicity during the epoch of formation of NGC 121 is consistent with the metallicity measured for NGC 121, which is also something not surprising.  Comparing with literature values, we note that the enrichment history is consistent with the cluster enrichment history given by \citet{ols96} and nearly identical to that measured by \citet{dac98} and calculated by \citet{pag98}.  The metallicity for old stars also agrees with that measured by \citet{sun86}.

\section{Summary}

We have presented WFPC2 $VI$ photometry of NGC 121 and a nearby outer SMC field.  The CMD of NGC 121 showed sharp features, as is expected for a single-population object.  Using a numerical fit of \citet{gir00} isochrones and observational errors to the observed CMD, we have measured a self-consistent solution of $\mu_0 = 18.96 \pm 0.04$ ($d = 61.9 \pm 1.1$ kpc), $A_V = 0.04 \pm 0.04$, $t = 10.6 \pm 0.5$ Gyr, $\logz = -1.03 \pm 0.06$, and $M_0 = 4.1 \pm 0.4 \times 10^5 M_{\odot}$ (again assuming a Salpeter IMF cutting off at $0.1 M_{\odot}$), with uncertainties accounting for random errors and possible calibration errors, but not possible systematic errors in the isochrones.  These values are all consistent with values measured through a variety of independent techniques -- dust maps of \citet{sch98} (extinction), $V_{HB}$ (distance), RGB color ($\logz$), and $\Delta V^{TO}_{HB}$ (age) -- as well as with previous measurements of \citet{str85} and \citet{mig98}.

The SMC field observations were taken $2\deg$ away from NGC 121 to avoid contamination from NGC 121 and from the nearby Galactic cluster 47 Tuc.  The primary analysis was done using ground-based data containing 16110 stars.  We also used four WFPC2 pointings in the same area, containing a total of 4141 stars, for a consistency check.  A qualitative examination of the resulting CMDs shows a very broad turnoff region, with stars at least 10 Gyr old and stars as young as 2 Gyr.

We made a numerical fit to the observed CMD in a similar manner as that made for NGC 121.  Combining our ground-based and WFPC2 data, we measured a distance and extinction of $\mu_0 = 18.88 \pm 0.08$ and $A_V = 0.05 \pm 0.07$, respectively, for this region of the SMC.  We found evidence of continuous star formation from ancient times until 2 Gyr ago, with an average star formation rate of $3.2 \pm 0.2 \times 10^{-4} M_{\odot} kpc^{-2} yr^{-1}$ (assuming a Salpeter IMF cutting off at $0.1 M_{\odot}$).  At the $1 \sigma$ level, we detect a broadly-peaked past star formation rate, with the maximum between $5-8$ Gyr ago.  At the $2 \sigma$ level, it is possible that the star formation rate was constant from 15 Gyr ago until 2 Gyr ago.  The oldest measurable star formation was $9-12$ Gyr ago; we can neither confirm nor rule out the presence of older stars.  Regardless of what one decides for the intermediate-age and ancient star formation history, the greatly decreased star formation in the past 2 Gyr (by an order of magnitude) is certain.

Our star formation history solution also derived a chemical enrichment history of the SMC.  We find a metallicity of $\logz = -1.3 \pm 0.3$ for the oldest stars, increasing to a value of $\logz = -0.7 \pm 0.2$ between $1-2$ Gyr ago, consistent with the findings of previous studies \citep{ols96,dac98,pag98}.  The lack of young stars prohibits a definitive measurement of the metallicities of younger stars.

\acknowledgments

Support for this work was provided by NASA through grant number GO-06604 from the Space Telescope Science Institute, which is operated by AURA, Inc., under NASA contract NAS 5-26555.

\clearpage

\begin{figure}
\caption{Cosmic ray cleaned deep F814W image of NGC 121. North is to the left and east is down.  (Image is given as a GIF file in preprint to meet the size limitations.)}
\label{fig_ngc121image}
\end{figure}
\clearpage

\begin{figure}
\caption{Color-magnitude diagrams of NGC 121. Panel (a) shows all stars located by \textit{multiphot}. Panel (b) shows all stars with good photometry ($\chi \le 2.0$ and $|\mbox{sharpness}| \le 0.3$). Panels (c) and (d) show color-magnitude diagrams of the good stars in the PC and WFC cameras, respectively.  (Image is given as a GIF file in preprint to meet the size limitations.)}
\label{fig_ngc121cmd}
\end{figure}
\clearpage

\clearpage

\clearpage

\clearpage

\clearpage

\clearpage

\clearpage

\clearpage

\clearpage

\clearpage

\clearpage

\clearpage

\clearpage

\begin{deluxetable}{lrr}
\tablecaption{HST WFPC2 Observations. \label{tab_fields}}
\tablewidth{0pt}
\tablehead{
\colhead{Pointing} &
\colhead{RA\tablenotemark{a}} &
\colhead{Dec\tablenotemark{a}}}
\startdata
NGC 121 & $00 26 42$ & $-71 31 00$ \\
Field-1 & $00 46 00$ & $-70 35 00$ \\
Field-2 & $00 49 00$ & $-70 48 00$ \\
Field-3 & $00 46 12$ & $-70 47 00$ \\
Field-4 & $00 49 00$ & $-70 33 00$ \\
\enddata
\tablenotetext{a}{All coordinates given in J2000.0}

\end{deluxetable}

\clearpage
\begin{deluxetable}{ccc}
\tablecaption{Field Star Formation History. \label{tab_fieldsfh}}
\tablewidth{0pt}
\tablehead{
\colhead{Age (Gyr)} &
\colhead{Mean SFR\tablenotemark{a}} &
\colhead{Mean $\logz$\tablenotemark{b}}}
\startdata
$ 0.0 -  0.5$ & $0.01 \pm 0.01$ & $-0.7 \pm 0.4$ \\
$ 0.5 -  1.0$ & $0.07 \pm 0.03$ & $-0.7 \pm 0.3$ \\
$ 1.0 -  2.0$ & $0.11 \pm 0.02$ & $-0.7 \pm 0.2$ \\
$ 2.0 -  3.0$ & $0.63 \pm 0.29$ & $-0.7 \pm 0.2$ \\
$ 3.0 -  5.0$ & $1.19 \pm 0.32$ & $-0.8 \pm 0.2$ \\
$ 5.0 -  8.0$ & $2.05 \pm 0.44$ & $-1.2 \pm 0.4$ \\
$ 8.0 - 11.0$ & $1.19 \pm 0.38$ & $-1.3 \pm 0.3$ \\
$11.0 - 15.0$ & $0.53 \pm 0.18$ & $-1.4 \pm 0.2$ \\
\enddata
\tablenotetext{a}{Mean star formation rates are calculated only from the ground-based observations, because of the larger field of view, and are normalized to the lifetime average value.}
\tablenotetext{b}{Mean metallicity measurements are made from both the ground-based and WFPC2 data.}

\end{deluxetable}

\end{document}